\documentclass[prper,aps,twocolumn,groupedaddress,floats,showpacs,final,superscriptaddress]{revtex4-2}
\usepackage[latin3]{inputenc}
\usepackage[makeroom]{cancel}
\usepackage{graphicx}
\usepackage{amsmath}
\usepackage{amsfonts}
\usepackage{amssymb}
\usepackage{chngcntr}
\usepackage{color}
\usepackage[left=2cm,right=2cm,top=2cm,bottom=2cm]{geometry}

\usepackage{graphicx}
\usepackage{dcolumn}
\usepackage{bm}
\usepackage{simplewick}
\usepackage{array}
\usepackage{appendix}

\RequirePackage[
   hyperindex,colorlinks,bookmarksnumbered,
   plainpages=true,pdfstartview=FitH]{hyperref}
\hypersetup{linkcolor=blue,urlcolor=blue,citecolor=blue}
\usepackage{hyperref}

\definecolor{purple}{rgb}{0.5,0,0.6}

\begin{document}

\title{Photogalvanic transport in fluctuating Ising superconductors}


\author{A.~V.~Parafilo}
\affiliation{Center for Theoretical Physics of Complex Systems, Institute for Basic Science (IBS), Daejeon 34126, Korea}

\author{M.~V.~Boev}
\affiliation{Novosibirsk State Technical University, Novosibirsk 630073, Russia}
\affiliation{Rzhanov Institute of Semiconductor Physics, Siberian Branch of Russian Academy of Sciences, Novosibirsk 630090, Russia}

\author{V.~M.~Kovalev}
\affiliation{Novosibirsk State Technical University, Novosibirsk 630073, Russia}

\author{I.~G.~Savenko}
\affiliation{Center for Theoretical Physics of Complex Systems, Institute for Basic Science (IBS), Daejeon 34126, Korea}
\affiliation{Basic Science Program, Korea University of Science and Technology (UST), Daejeon 34113, Korea}


\date{\today}

\begin{abstract}
In a two-dimensional noncentrosymmetric Ising superconductor in the fluctuating regime under the action of a uniform external electromagnetic field there emerge two contributions to the photogalvanic effect due to the trigonal warping of the valleys. The first contribution stems from the current of the electron gas in its normal state, while the second contribution is of Aslamazov-Larkin nature: it originates from the presence of fluctuating Cooper pairs when the ambient temperature approaches (from above) the temperature of superconducting transition in the sample. 
The way to lift the valley degeneracy is the application of a weak out-of-plane external magnetic field producing a Zeeman effect. The Boltzmann equations approach for the electron gas in the normal state and the time-dependent Ginzburg-Landau equations for the fluctuating Cooper pairs allow for the study of the photogalvanic current in two-dimensional transition metal dichalcogenide Ising superconductors.  
\end{abstract}

\maketitle

\section{Introduction}\label{intro}

In two-dimensional (2D) materials, photoinduced transport phenomena, which are second-order with respect to external electromagnetic (EM) field, are in the focus of state-of-the-art research~\cite{glazovganichev}. 
The majority of these effects fall into two categories. 
The first one includes the rectification effects to occur under an external uniform alternating EM illumination, which produces stationary uniform electric currents in the system. The second category encompasses all the effects characterised by the system response at doubled field frequency, thus describing the second-harmonic generation phenomena. 

The second-order transport phenomena are usually sensitive to the polarization of the EM field and the symmetry of the system under study, namely, the time-reversal symmetry and the spatial inversion symmetry. 
The phenomenological relation between the photoinduced rectified electric current and the amplitude of external EM field reads $j_\alpha=\zeta_{\alpha\beta\gamma}E_\beta E^*_\gamma$, where $\zeta_{\alpha\beta\gamma}$ is the third-order tensor acquiring non-zero components in non-centrosymmetric materials. 
In non-gyrotropic semiconductor materials, the (rectified) photoinduced  electric current occurs as a second-order response to linearly polarized external EM wave.
This constitutes the photogalvanic effect (PGE). 
This effect does not directly relate to either light pressure, the photon-drag phenomena, or non-uniformity of either the sample or light field intensity, like the photoinduced Dember effect. 
Instead, the microscopic origin of the conventional PGE lies in the asymmetry of the interaction potential or the crystal-induced Bloch wave function~\cite{sturman, ivchenko, ganichev}. 

In modern Wan-der-Waals structures based on 2D monolayers of transition-metal dichalcogenides (TMDs)~\cite{wang2012electronics,manzeli20172d}, 
the PGE current may arise due to specific band structure of the material possessing two time-reversal-coupled valleys in the Brillouin zone. 
A typical example of these materials is molybdenum disulfide, MoS$_2$. 
It possesses the D$_{3h}$ point group, and the presence of the $C_3$ axis results in the emergence of a trigonal warping of the electron dispersion in each valley reflecting the noncentrsymmetricity of the crystal structure. 
The theoretical analysis shows that the PGE current arises here in each valley (involving electrons residing in both the valleys), and these currents have different signs in different valleys. 
As a result, net PGE current self-compensates and vanishes.
A nonzero net current may only occur if the time-reversal symmetry is broken due to, e.g., the presence of external magnetic field or illumination of the sample by a circularly-polarized EM field causing interband transitions~\cite{Entin, kovalev, Entin2}.   

Furthermore, a resent discovery of the superconducting (SC) transition in TMDs~\cite{MoS2superconductivity, doi:10.1126/science.aab2277, CostanzoNat} stimulated additional interest to the study of transport phenomena in 2D Dirac materials exposed to external EM fields at lower temperatures~\cite{Sun_20212DMater, PhysRevB.99.115408, PhysRevLett.124.087701}.
In the intermediate range of temperatures lying in between the normal and SC state of the electron gas, when $0<T-T_c\ll T_c$ (where $T_c$ is a SC critical temperature), the order parameter starts to experience fluctuations~\cite{AL,Varlamov, PhysRevLett.124.207002}. 
Moreover, large spin-orbit coupling in TMDs results in strong out-of-plane electron spin polarization and large in-plane critical magnetic fields beyond the Pauli limit. 
Thus, all the ingredients of an \textit{Ising superconductor} possessing unique physical properties are available. 
When the time-reversal symmetry breaks by a weak  magnetic field due to the Zeeman effect, and given the absence of spatial reversal symmetry, TMDs might demonstrate pronounced nonreciprocal response in the regime of SC fluctuations~\cite{ doi:10.1126/science.aab2277, CostanzoNat, SaitoNat, XiNat, PhysRevLett.113.097001, LI2021100504}.


The goal of this work is to develop a microscopic theory of a linear PGE effect in fluctuating Ising superconductors exposed to linearly-polarized EM field. 
The time-reversal symmetry here is waved due to the presence of a weak Zeeman field pointed across the monolayer \cite{nagaosa,nagaosa2}.
Trigonal warping of the valleys $K$ and $K'$ characteristic of MoS$_2$ serves as a microscopic mechanism of the effect.
Within the D$_{3h}$ point symmetry group, the third-order conductivity (or transport coefficient) tensor possesses only one nonzero component. 
Thus, phenomenologically, the PGE current can be expressed as $j_x=\zeta(|E_x|^2-|E_y|^2),\,\,j_y=-\zeta(E_xE^*_y+E_x^*E_y)$. 
Therefore, the main task comes down to the calculation of the coefficient $\zeta$ and analyzing its behavior for various EM field frequencies and temperatures in the vicinity of $T_c$, taking into account the contribution of normal electrons and the corrections arising from the SC order parameter fluctuations. 

\section{Effective electron dispersion in  conduction band} \label{spectrum}


The superconducting transition in MoS$_2$ monolayer occurs at electron densities exceeding $10^{14}$~cm$^{-2}$ \cite{MoS2superconductivity}. At such high densities, the Fermi level lies deeply in the conduction band. 
Thus, it is feasible to use a simplified electron energy dispersion. 
Then, according to the two-band model, the Hamiltonian reads (in $\hbar=k_B=1$ units)
\begin{gather}\label{Zeeman1}
H=\frac{\Delta}{2}\sigma_z+v(\eta p_x\sigma_x+p_y\sigma_y)+\left(
                                                            \begin{array}{cc}
                                                              0 & \mu p_+^2 \\
                                                              \mu p_-^2 & 0 \\
                                                            \end{array}
                                                          \right)\\\nonumber
+s\eta\frac{\lambda_c}{2}(\sigma_z+1)
-s\eta\frac{\lambda_v}{2}(\sigma_z-1)+s\Delta_Z,
\end{gather}
where $\Delta$ is the material bandgap, $\sigma_i$ are the Pauli matrices, $v$ is the band parameter with the dimensionality of velocity, $\eta=\pm1$ is the valley index, $\mathbf{p}$ is the electron momentum,  $p_{\pm}=\eta p_x\pm i p_y$, $\mu$ is the band parameter describing the trigonal warping and nonparabolicity of electron dispersion, $s$ is the $z$-component of electron spin, $\lambda_{c,v}$ describe spin-orbit splitting of the conduction and valence bands, and $\Delta_Z\propto B$ is the Zeeman energy due to the external magnetic field applied across the monolayer plane.

The eigenvalues of the Hamiltonian~\eqref{Zeeman1} read 
\begin{eqnarray}\label{Zeeman2}
&&E_{s\eta}({\bf p})=s\Delta_Z+s\eta\frac{\lambda_c+\lambda_v}{2}
\\\nonumber
&&~~~~~~~~~~~~~~\pm\sqrt{\left[\frac{\Delta-s\eta(\lambda_v-\lambda_c)}{2}\right]^2+|h_{\bf p}|^2},\\
\nonumber
&&~~~~~\frac{|h_{\bf p}|^2}{\Delta}=\epsilon_p+\eta w(p_x^3-3p_xp_y^2),
\end{eqnarray}
where $\epsilon_p=p^2/2m$ is electron kinetic energy in the conduction band, $m=\Delta/(2v^2)$ is the electron effective mass, $w=2v\mu/\Delta$ is a warping amplitude. 
Expression~\eqref{Zeeman2} can be further expanded using the inequality $|h_{\bf p}|^2/\Delta^2\ll1$, 
\begin{gather}\label{cond_band}
E_{s\eta}({\bf p})
\approx s(\Delta_Z+\eta\lambda_c)
+\epsilon_p+\eta w(p_x^3-3p_xp_y^2),
\end{gather}
counting the conduction band energy from the value $\Delta/2$.


\section{Normal-state electron gas  contribution to PGE}\label{normal}

Let us, first, study the PGE current of normal-state electrons exposed to a uniform external EM field ${\bf E}(t)={\bf E}e^{-i\omega t}+{\bf E}^*e^{i\omega t}$ with normal incidence to the monolayer, thus ${\bf E}=(E_x,E_y,0)$.
In the case $\omega\ll\epsilon_F$, where $\epsilon_F$ is the Fermi energy, the Boltzmann equation~\cite{Zaitsev:2014aa, Ziman:2001aa} represents a suitable tool to analyse the PGE transport~\cite{glazovganichev, Entin}. 
In the framework of the (single) relaxation time  approximation, the Boltzmann equation reads
\begin{gather}\label{Zeeman3}
\frac{\partial f}{\partial t}+e{\bf E}(t)\cdot\frac{\partial f}{\partial {\bf p}}=-\frac{f-f_0}{\tau},
\end{gather}
where $f$ is the electron distribution function, $f_0$ is the Fermi distribution, $e$ is the elementary charge, $\tau$ is the scattering time (on the point-like impurities). 
In the expansion $f=f_0+f_1(t)+f_2+f_2(t)+...$ with respect to the amplitude of external electric field, the first-order correction depends on time, $f_1(t)=f_1e^{-i \omega t}+f_1^*e^{i \omega t}$, whereas the second-order correction consists of the stationary, $f_2$, and alternating  part, $f_2(t)$. 
Summing up all the first-order terms yields
\begin{gather}\label{Zeeman4}
f_1=-e\tau_{\omega}{\bf E}\cdot\frac{\partial f_0}{\partial {\bf p}}=e\tau_{\omega}{\bf v}\cdot {\bf E}(-f_0'),
\end{gather}
where $\tau_\omega=\tau/(1-i\omega\tau)$, $f_0'=\partial f_0/\partial E_{s\eta}$, and the electron velocity reads ${\bf v}=\partial_{\bf p}E_{s\eta}({\bf p})$. 

The stationary part of the second-order correction reads
\begin{gather}\label{Zeeman5}
f_2=-e\tau\left({\bf E}\cdot\frac{\partial f_1^*}{\partial {\bf p}}+{\bf E}^*\cdot\frac{\partial f_1}{\partial {\bf p}}\right),
\end{gather}
which determines the PGE current,
\begin{gather}\label{Zeeman6}
j_\alpha=e\int\frac{d{\bf p}}{(2\pi)^2}v_\alpha\,f_2, ~~~~~\alpha=x,~y.
\end{gather}
Combining Eq.~\eqref{Zeeman4} and Eq.~\eqref{Zeeman5}, and integrating by parts in Eq.~\eqref{Zeeman6}, yields the expression for the PGE current density in the form,
\begin{eqnarray}\label{Zeeman7}
&&j_\alpha=e^3\tau
\left(\tau^*_\omega E_\beta E^*_\gamma+\tau_\omega E^*_\beta E_\gamma\right)
\\
\nonumber
&&~~~~~~~~~~\times
\sum_{s,\eta}\int\frac{d{\bf p}}{(2\pi)^2}\frac{\partial^2 v_\alpha}{\partial p_\beta\partial p_\gamma}\,f_0[E_{s\eta}({\bf p})];
\\
\nonumber
&&~~~~~~~~~~\frac{\partial^2 v_\alpha}{\partial p_\beta\partial p_\gamma}\equiv\frac{\partial^3 E_{s\eta}({\bf p})}{\partial p_\beta\partial p_\gamma\partial p_\alpha},
\end{eqnarray}
where $f_0[E_{s\eta}({\bf p})]=\theta[\epsilon_F-E_{s\eta}({\bf p})]$ for a degenerate electron gas with $\theta[x]$ the Heaviside step function. 
Accounting that $\partial^2v_{\alpha}/\partial p_{\beta}\partial p_{\gamma}= \pm 6\eta w $ (here, $"+"$ stands for $xxx$ component, while $"-"$ stands for $xyy$, $yxy$, $yyx$ components, the others are zero), from Eq.~(\ref{Zeeman7}) it follows that the PGE current is proportional to the differences between electron densities, $n_{\pm}$, in both valleys, $j_{\alpha}\propto \sum_{s,\eta} \eta \,n_{\eta}$. 
Therefore, the normal-state electron gas does not contribute to the nonreciprocal current in the framework of this model, as it is also claimed in work~\cite{nagaosa}. 
The reason for such behavior is that the Zeeman field only redistributes the electrons between spin-resolved subbands in each valley, keeping the total electron density in the valley unchanged.

In order to have a finite PGE response, it is necessary to modify the original model, Eqs.~(\ref{cond_band}) and~\eqref{Zeeman3}, by introducing energy-dependent relaxation time, $\tau_{\varepsilon}$. 
In a particular case of electron scattering on Coulomb impurities in a 2D system, the relaxation time is proportional to the electron energy, $\tau_{\varepsilon}=\tau_0\varepsilon_{\textbf{p}}$, where $\tau_0$ is a coefficient. 
Henceforth, in Eqs.~(\ref{Zeeman3})--(\ref{Zeeman6}), $\tau$ should be replaced by $\tau_{\varepsilon}=\tau_0\varepsilon_{\textbf{p}}=\tau_0[\epsilon_p+\eta w(p_x^3-3p_xp_y^2)]$. 
Then, the PGE current density in the static limit $\omega\tau_{\varepsilon}$$=$$\tau_0\omega\epsilon_F\ll 1$ (which additionally provides the relation between $\omega$ and the doping) reads
%
\begin{eqnarray}\label{current_rtime}
j_\alpha=2e^3 E_\beta E^*_\gamma
\sum_{s,\eta}\int\frac{d{\bf p}}{(2\pi)^2}v_{\alpha}\tau_{\varepsilon}\frac{\partial}{\partial p_{\beta}}\left\{\tau_{\varepsilon}\frac{\partial}{\partial p_{\gamma}} f_0\right\}.
\end{eqnarray}
%
To derive~\eqref{current_rtime}, we also assumed the absence of the inter-valley scattering~\cite{AntiPekka2019scattering} and neglected the spin-flip processes transferring the electrons between spin-resolved subbands in a given valley. 

Expanding Eq.~(\ref{current_rtime}) in the lowest-order in $w$ and restoring dimensionality yields (see the Supplemental Material~\cite{[{See Supplemental Material at [URL] for the detailed  derivations}]SMBG})
%
\begin{eqnarray}\label{normal_state_PGE}
\textbf{\textit{j}}=108\frac{e^3\tau_0^2\Delta_Z\lambda_c w n_e}{ \hbar^3}\textbf{F}(\textbf{E}),
\end{eqnarray}
where $\textbf{F}(\textbf{E})=(|E_x|^2-|E_y|^2,-E_xE_y^{\ast}-E_yE_x^{\ast})$, and $n_e=n_{+}+n_{-}$ is a total electron density in both the valleys. 
The PGE current of the normal-state electron gas, Eq.~(\ref{normal_state_PGE}) represents the first important result of this article: 
Nonreciprocal PGE response is finite in the case of electron scattering off Coulomb impurities in 2D samples. 

Taking the electron density $n_e\sim 10^{14}\,{\rm cm}^{-2}$, the external magnetic field $B=1\,\textrm{T}$, the amplitude of EM field $E_0=1\, \textrm{V/cm}$,  $\tau_0=10 \,\textrm{ps/eV}$ (which is the highest possible value found from the relation $\tau_0\omega \epsilon_F\ll1$ for $n_e=10^{14}\,\textrm{cm}^{-2}$ and $\omega=0.1\,\textrm{ps}^{-1}$), and typical parameters for MoS$_2$~\cite{Kormanyos, nagaosa}, $\lambda_c=3$~meV and $w=-3.4 \,\textrm{ eV}\cdot\textrm{\rm \AA}^3$, we find that a typical magnitude of the PGE current due to the normal 2D electron gas contribution amounts to $j\sim 10$~nA/cm.

\section{Superconducting fluctuations  contribution to PGE}\label{superconducting}

The electric current density operator due to the presence of SC fluctuations reads 
\begin{eqnarray}\label{current}
\textbf{\textit j}=\frac{e^{\ast}}{2}\left\{\Psi^{\ast} {\bf v}(\hat{\textbf{p}})\Psi+\Psi {\bf v}(-\hat{\textbf{p}})\Psi^{\ast}\right\},
\end{eqnarray}
where $e^{\ast}=2e$ is a charge of a Cooper pair, ${\bf v}(\hat{\textbf{p}})$ is a Cooper pair velocity operator, ${\hat {\bf p}}=-i\nabla$ is a momentum operator, and the superconducting order parameter $\Psi (\textbf{r}, t)$ satisfies the time-dependent Ginzburg-Landau (TDGL) equation with account of the trigonal warping contribution to the kinetic energy of a Cooper pair,
\begin{eqnarray}\label{TDGLequation}
\left[
\gamma\frac{\partial}{\partial t}
+\varepsilon(\hat{{\bf p}})+2ie\gamma \varphi(\textbf{r}, t)
\right]
\Psi(\textbf{r},t)=f(\textbf{r},t).
\end{eqnarray}
In Eq.~\eqref{TDGLequation}, $\gamma=\pi\alpha/8$, $\alpha$ is the parameter of GL theory, which is inversely proportional to the effective mass $m$ and square of the coherence length $\xi$: $4m\alpha T_c\xi^2=1$, thus,  $\varepsilon({\bf p})=p^2/4m +\alpha T_c\epsilon+\Lambda(p_x^3-3p_xp_y^2)\equiv\alpha T_c(\epsilon+p^2\xi^2)+\Lambda(p_x^3-3p_xp_y^2)$ is the Cooper pair kinetic energy, 
and $\epsilon=(T-T_c)/T_c$ is the reduced temperature.
The coherence length in 2D reads
\begin{eqnarray}\label{coherence_length}
\xi^2=\frac{v_F^2\tau^2}{2}\Bigl[
\psi\left(\frac{1}{2}\right)
-
\psi\left(\frac{1}{2}+\frac{1}{4\pi T\tau}\right)
+\frac{\psi'\left(\frac{1}{2}\right)}{4\pi T\tau}\Bigr],~~~
\end{eqnarray}
where $\psi(x)$ is the digamma function, and $v_F=\sqrt{4\pi n_e}/m$ is the Fermi velocity.
Furthermore, in Eq.~\eqref{TDGLequation}, $\varphi(\textbf{r}, t)=\varphi e^{i\textbf{k}\textbf{r}-i\omega t}+\varphi^{\ast} e^{-i\textbf{k}\textbf{r}+i\omega t}$ is the scalar potential, which obeys standard correspondence with the external uniform EM field, $\textbf{E}=-\nabla\varphi$.

The Cooper pair trigonal warping amplitude, $\Lambda$, entering Eq.~\eqref{TDGLequation} through the term $\varepsilon(\hat{{\bf p}})$, in a clean superconductor ($\tau T_c\gg1$) and for the $s$-wave singlet pairing can be expressed through the Zeeman field and the normal electrons warping amplitude $w$~\cite{nagaosa},
\begin{gather}\label{tdgl3}
\Lambda=\frac{93\zeta(5)\Delta_Z\lambda_cw}{28\zeta(3)(\pi T_c)^2}.
\end{gather}
To estimate it, let us substitute typical parameters for MoS$_2$ (given in the last paragraph of the previous section) 
and the SC critical temperature $T_c=10\,\textrm{K}$: $|\Lambda|\approx 0.46 \,\textrm{eV}\cdot\textrm{\AA}^3$ for  $B=1\,\textrm{T}$.

The r.h.s. of Eq.~\eqref{TDGLequation} is the Langevin force, describing SC fluctuations in the equilibrium.
It satisfies the white-noise law,
\begin{eqnarray}\label{noise}
\langle f^{\ast}(\textbf{r})f(\textbf{r}',t')\rangle= 2\gamma T \delta(\textbf{r}-\textbf{r}')\delta(t-t'),
\end{eqnarray}
which allows us to find an expression for the SC order parameter in equilibrium, $\langle |\Psi_{0\textbf{p}}|^2\rangle=[\alpha(\epsilon+\xi^2\textbf{p}^2)]^{-1}$ (here, $\Psi_{0\textbf{p}}$ is the Fourier transform of the order parameter).

As concerns the applicability of the TDGL equation in the form~\eqref{TDGLequation}, it is only valid in the low-frequency domain ($\omega\tau\ll1$) given an arbitrary ration between $\omega$ and $T_c$.  
In the range of moderate and high frequencies ($\omega\tau\gtrsim 1$), various non-locality corrections emerge~\cite{PhysRevB.51.3880}.
Treating them requires the usage of quantum-field theory approaches beyond the TDGL equation. 
Therefore, this theory is applicable to either clean superconductors, $\omega<\tau^{-1}<T_c$, or ``dirty'' superconductors obeying the relation $(\omega,T_c)\tau<1$.
Moreover, in addition to Aslamazov-Larkin correction there exist other fluctuating contributions, such as the Maki-Tompson~\cite{Maki,Tompson} and the ``density of states''~\cite{dos} ones, which are beyond the scope of present paper. 
Treating them also requires the using of quantum-field theory approaches beyond the TDGL equation~\cite{Varlamov}.

Let us start with a clean superconductor case.
Expanding the order parameter with respect to the scalar potential, $\Psi(\textbf{r},t)=\Psi_{0}(\textbf{r},t)+\Psi_{1}(\textbf{r},t)+\Psi_{2}(\textbf{r},t)+...$, and then, substituting this expansion in Eq.~\eqref{current} keeping only the second-order terms, gives two contributions to the electric current density,
\begin{eqnarray}
 j^{I}_{\alpha}=&&e\left\{\Psi_{1}^{\ast}v_{\alpha}(\hat{\textbf{p}})\Psi_{1}+\Psi_{1} v_{\alpha}(-\hat{\textbf{p}})\Psi_{1}^{\ast}
\right\},\label{current1}\\
 j^{II}_{\alpha}=&&e\left\{\Psi_{0}^{\ast}v_{\alpha}(\hat{\textbf{p}})\Psi_{2}+\Psi_{2}^{\ast}v_{\alpha}(\hat{\textbf{p}})\Psi_{0}\right\}\nonumber\\+&&e\left\{\Psi_{0}^{\ast}v_{\alpha}(-\hat{\textbf{p}})\Psi_{2}+\Psi_{2}^{\ast}v_{\alpha}(-\hat{\textbf{p}})\Psi_{0}\right\},\label{current2}
\end{eqnarray}
%
where
\begin{eqnarray}
&&\Psi_{0}(x)=\int dx' g(x-x')f(x'),\label{psi1}\\
&&\Psi_{1(2)}(x)=-2ie\gamma\int dx' g(x-x')\varphi(x')\Psi_{0(1)}(x').~~~\label{psi2}
\end{eqnarray}
with $x=(\textbf{r},t)$ the short-hand notation, and
\begin{eqnarray}
\label{EqFluctProp}
g(\textbf{r},t)=\sum_{\varepsilon,\textbf{p}}e^{i\textbf{p}\textbf{r}-i\varepsilon t}g_{\textbf{p}}(\varepsilon)\equiv\sum_{\varepsilon,\textbf{p}}\frac{ e^{i\textbf{p}\textbf{r}-i\varepsilon t}}{-i\gamma \varepsilon +\varepsilon(\textbf{p})}
\end{eqnarray}
the fluctuation propagator in standard form.

Combining Eqs.~\eqref{current1}--\eqref{EqFluctProp} and performing the averaging over the fluctuating Langevin forces gives a general expression for the PGE current:
%
\begin{eqnarray}
\nonumber
&&j_{\alpha}=2(2e\gamma)^3T\sum_{\varepsilon,\textbf{q}}  
|g_{\textbf{q}}(\varepsilon)|^2 
\left\{
|g_{\textbf{k+q}}(\varepsilon+\omega)|^2  v_{\alpha}(\textbf{q}+\textbf{k})
\right.
\\
\nonumber
&&\left.~~- v_{\alpha}(\textbf{q})[ g_{\textbf{q}}(\varepsilon)g_{\textbf{q}+\textbf{k}}(\varepsilon+\omega)+g_{\textbf{q}}^{\ast}(\varepsilon)g_{\textbf{q}+\textbf{k}}^{\ast}(\varepsilon+\omega)]
\right\} 
|\varphi|^2 
\\
\nonumber
\\
\label{pge}
&&~~~~~~~~~~~~~~~~~+(\textbf{k}\rightarrow -\textbf{k},\, \omega \rightarrow -\omega).\quad
\end{eqnarray}
%
Let us mention, that in Eq.~\eqref{pge}, only the static contribution to the product of two scalar potentials is accounted for, thus disregarding the $2\omega$ harmonics. 

After the integration over energy, Eq.~\eqref{pge} acquires a more compact form, 
%
\begin{eqnarray}
\label{fluc_current}
&&j_{\alpha}=(2e)^3\gamma^2T\sum_{\textbf{q}}   \left\{\frac{v_{\alpha}(\textbf{q}+\textbf{k})}{\varepsilon(\textbf{q})\varepsilon(\textbf{q}+\textbf{k})} -\frac{ v_{\alpha}(\textbf{q})}{\varepsilon^2(\textbf{q})}\right\}\\
\nonumber
&&~~~~~~~~~\times\frac{\varepsilon(\textbf{q})+\varepsilon(\textbf{q}+\textbf{k})}{\gamma^2\omega^2+\left[\varepsilon(\textbf{q})+\varepsilon(\textbf{q}+\textbf{k})\right]^2}|\varphi|^2\\
\nonumber
&&\\
\nonumber
&&~~~~~~~~~~~~~
+(\textbf{k}\rightarrow -\textbf{k}, \omega \rightarrow -\omega).
\end{eqnarray}
%
Evidently, this current vanishes at $\textbf{k}\rightarrow 0$. 
It motivates the need to expand the PGE current up to the second order over $\textbf{k}$ using the correspondence between the electrostatic potential and components of the electric field,  $(-ik_{\beta})(ik_{\gamma})|\varphi|^2=E_{\beta}E^{\ast}_{\gamma}$. 
The first-order corrections vanish since the terms with opposite signs (directions) of $\textbf{k}$ cancel each other out.

Furthermore, expanding $\varepsilon(\textbf{p})$ in Eq.~\eqref{fluc_current} 
up to the first order in warping $\Lambda$ and integrating over the momentum $\textbf{q}$, gives the paraconductivity contribution to the PGE as $\textbf{\textit{j}}=\zeta_S\mathbf{F}$, where after restoring dimensionality
\begin{eqnarray}\label{PGE_current}
&&\zeta_S=\frac{3e^3\Lambda m \pi}{16\hbar^3k_B T_c \epsilon^2}\frac{1}{\tilde\omega^2}\left[1+\frac{\log\left(1+\tilde\omega^2\right)}{\tilde\omega^2}+\frac{\pi}{2} \left(\frac{1}{\tilde\omega^3}-\frac{1}{\tilde\omega}\right)\right.\nonumber\\
&&\left.-\frac{1}{\tilde\omega}\left(1+\frac{1}{\tilde\omega^2}\right)\arctan\tilde\omega+\frac{1}{\tilde\omega}\left(1-\frac{1}{\tilde\omega^2}\right)\arctan\frac{1}{\tilde\omega}\right],\\
\nonumber
&&
\\
\nonumber
&&~~~~~~~~~~~~~~~~~~~~~~~\tilde\omega=\frac{\pi\hbar\omega}{16k_B(T-T_c)},
\end{eqnarray}
which represents the second important result of this paper.

The third-order {\it ac} paraconductivity tensor $\zeta_S$ experiences its maximum at the static limit, $\zeta_S(\tilde\omega\ll 1)\approx\zeta_0(1-2\tilde\omega^2/5)$, whereas it decays with the increase of the frequency of the EM field as $\zeta_S(\tilde\omega\gg 1)\approx 6\zeta_0/\tilde\omega^2$, where $\zeta_0=2e^3\Lambda m\pi /64\hbar^3T_c\epsilon^2$ is the paraconductivity tensor for the {\it dc} nonreciprocal current~\cite{nagaosa,nagaosa2} (interestingly, using the Boltzmann kinetic equation gives the same result, see the Supplemental Material~\cite{[{See Supplemental Material at [URL] for the detailed  derivations}]SMBG}).

Evidently, the dc component of $\zeta_S$ decays as $(T-T_c)^{-2}$ that is much faster than the conventional Aslamazov-Larkin correction in 2D, $\sigma^{AL}\propto (T-T_c)^{-1}$~\cite{AL}. Moreover, the paraconductivity starts to decrease rapidly with the increase of frequency even for $\omega\gtrsim(T-T_c)$ while the power of this decrease coincides with the power of $\epsilon$-dependence of the dc conductivity component. 
Note, Eq.~(\ref{PGE_current}) is only valid for a linearly polarized EM field, while the PGE vanishes in the case of a circularly polarized light. 

The next task is to generalize Eq.~(\ref{PGE_current}) for the case of an arbitrary impurity concentration by accounting for the relaxation time in the derivation of the Ginzburg-Landau free energy using the Green's function technique, see the Supplemental Material~\cite{[{See Supplemental Material at [URL] for the detailed  derivations}]SMBG}.
Indeed, the influence of SC fluctuations might be more prominent in dirty samples in accordance with the Ginzburg-Levanyuk criterion~\cite{levanyuk1959contribution}. The calculations in the case of an arbitrary $\tau T_c$ show that instead of the trigonal warping amplitude for the Cooper pairs $\Lambda$, which enters Eq.~\eqref{PGE_current} in the clean limit, there comes in play an effective warping coefficient, $\Lambda_{\tau}=\Lambda\cdot f_{\tau}(2\pi T_c \tau)$, where
%
%
\begin{eqnarray}
\nonumber
f_{\tau}(x)&=&\frac{7\zeta(3)}{31\zeta(5)}\frac{\pi n_e}{m^2}\frac{x^3}{(\pi T_c \xi)^2 }\left\{-2\pi^2+4\psi'\left(\frac{1}{2}+\frac{1}{2x}\right)\right.\\
\label{dirty_coeff}
&&\left.+\frac{1}{x}\left[14\zeta(3)-\psi''\left(\frac{1}{2}+\frac{1}{2x}\right)\right]\right\},
\end{eqnarray}
%
%
which represents a monotonous function of $\tau$, and $f_{\tau}(2\pi T_c\tau \gg 1)\rightarrow1$ in the  limit of a clean superconductor, whereas it vanishes linearly in the dirty case,  $f_{\tau}(2\pi T_c\tau \ll 1)\rightarrow0$. 
Interestingly enough, the trigonal warping term in the Ginzburg-Landau free energy and, as the consequence, in the photogalvanic current depends on the coherence length $\xi$ and the relaxation time $\tau$. 
Thus, the Cooper pairs in the fluctuating regime turn out sensitive to the presence of impurities in the sample, which is in contrast with the conclusions of the Aslamazov-Larkin theory being applied to the first-order response current. 
\begin{figure}[t!]
\includegraphics[width=0.9\columnwidth]{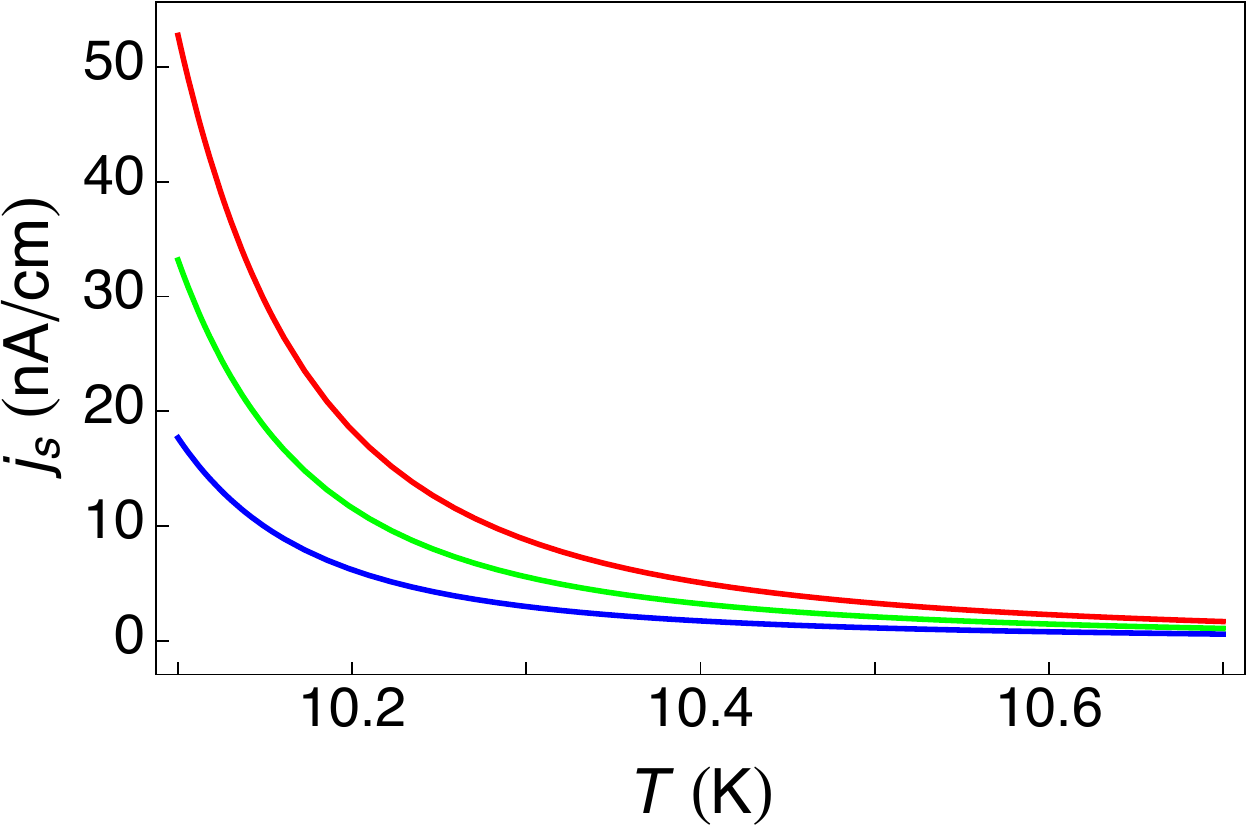}
\caption{Photogalvanic current of fluctuating Cooper pairs  [Eq.~(\ref{PGE_current}) where $\Lambda$ is replaced with $\Lambda_{\tau}$] as a function of temperature for $\omega=0.1$~ps$^{-1}$ and various relaxation times: $\tau=5\,{\rm ps}$ (red), $\tau=0.3\,{\rm ps}$ (green), and $\tau=0.1\,{\rm ps}$ (blue). All parameters are for MoS$_2$: superconducting critical temperature $T_c=10\,{\rm K}$, the warping amplitude $w=3.4\,{\rm eV}\cdot\textrm{\AA}^3$, $\lambda_c=3\,{\rm meV}$, $B=1 \,{\rm T}$, and the amplitude of electromagnetic field is $E_0=1\, {\rm V/cm}$.}
\label{Fig1}
\end{figure}

\begin{figure}[t!]
\includegraphics[width=0.9\columnwidth]{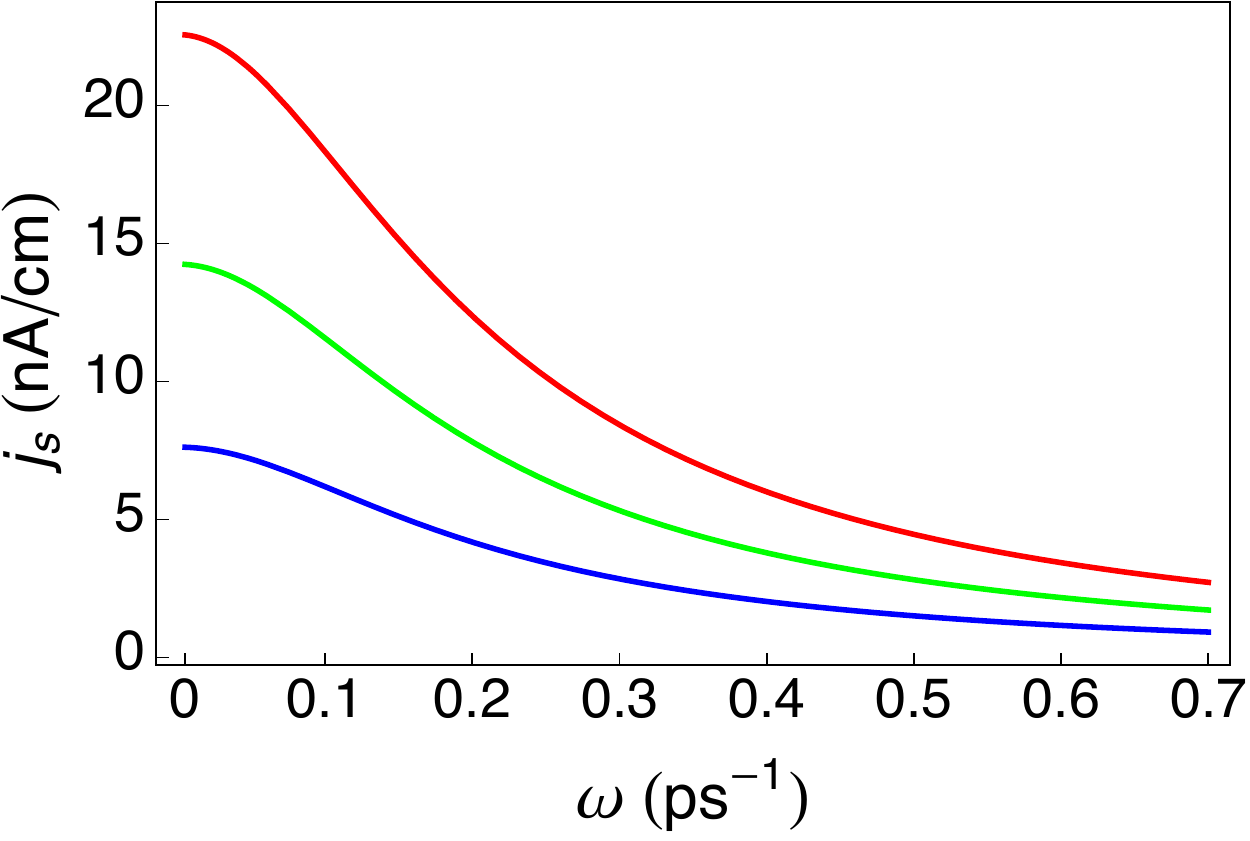}
\caption{Photogalvanic current of fluctuating Cooper pairs [Eq.~(\ref{PGE_current}) where $\Lambda$ is replaced with $\Lambda_{\tau}$] as a function of electromagnetic field frequency for $T=10.2\,\textrm{K}$ and various relaxation times: $\tau=5\,{\rm ps}$ (red), $\tau=0.3\,{\rm ps}$ (green), and $\tau=0.1\,{\rm ps}$ (blue). All other parameters are the same as in Fig.~\ref{Fig1}.}
\label{Fig2}
\end{figure}

Figure~\ref{Fig1} and Fig.~\ref{Fig2} show the temperature and frequency dependencies of the PGE current. 
Red curves correspond to the case of a clean superconductor, $\tau T_c\gg 1$. 
In terms of the EM field intensity, $I=c\epsilon_0|\textbf{E}|^2/2$ with $c$ the speed of light and $\epsilon_0$ the vacuum permittivity, the estimation gives $j/I\approx 4\,\textrm{nA}\cdot\textrm{cm}/\textrm{W}$ for $T=10.1\,\textrm{K}$ and $B=1\,\textrm{T}$.
Green and blue curves correspond to the case of dirty superconductors, $\tau T_c\ll 1$ and demonstrate the effect of the point-like impurities on the temperature and frequency dependencies of the PGE contribution due to SC fluctuations.


%
%
%
%
%
%


\section*{Conclusions}\label{conclusion}
We conclude, that in a two-dimensional noncentrosymmetric fluctuating Ising superconductor possessing trigonal warping of the valleys and exposed to a uniform external electromagnetic field there emerge two contributions to the photogalvanic effect.
The first contribution originates from the normal-state electron gas in the presence of Coulomb impurities in the sample.
The second contribution stems from the presence of superconducting fluctuations. 
In order to lift the valley degeneracy and, thus, have a nonzero photogalvanic electric current in the system, it is sufficient to use a weak out-of-plane external magnetic field producing a Zeeman effect and breaking the time-reversal symmetry.

The photogalvanic effect, thus, possesses Aslamazov-Larkin nature since it originates from the presence of fluctuating Cooper pairs when the ambient temperature approaches the temperature of superconducting transition in the sample. 
The electric current, as a second-order response of the system, possesses, first, a more pronounced temperature divergence $(T-T_c)^{-2}$, as compared with the Aslamazov-Larkin correction to the Drude conductivity, and, second, the current density has no smallness related to the electron-hole asymmetry of the quasiparticle spectrum, as it takes place in other second-order response effects~\cite{PhysRevB.52.7516, PhysRevLett.126.137002, PhysRevB.101.104512}.

\section*{Acknowledgement}
We were supported by the Institute for Basic Science in Korea (Project No.~IBS-R024-D1), Ministry of Science and Higher
Education of the Russian Federation (Project FSUN-2020-0004), and the Foundation for the Advancement of Theoretical Physics and Mathematics ``BASIS''.

\bibliography{biblio}
\bibliographystyle{apsrev4-2}



\end{document}


\title{Supplemental Material: Photogalvanic Transport \\
in Fluctuating Ising Superconductors}


\author{A.~V.~Parafilo}
\affiliation{Center for Theoretical Physics of Complex Systems, Institute for Basic Science (IBS), Daejeon 34126, Korea}

\author{M.~V.~Boev}
\affiliation{Novosibirsk State Technical University, Novosibirsk 630073, Russia}
\affiliation{Rzhanov Institute of Semiconductor Physics, Siberian Branch of Russian Academy of Sciences, Novosibirsk 630090, Russia}

\author{V.~M.~Kovalev}
\affiliation{Novosibirsk State Technical University, Novosibirsk 630073, Russia}

\author{I.~G.~Savenko}
\affiliation{Center for Theoretical Physics of Complex Systems, Institute for Basic Science (IBS), Daejeon 34126, Korea}
\affiliation{Basic Science Program, Korea University of Science and Technology (UST), Daejeon 34113, Korea}

\date{\today}

\maketitle

\tableofcontents
\vspace{20pt}

\section{Normal-state electron gas contribution to the PGE}

The Hamiltonian of electrons in MoS$_2$ reads as (in $\hbar=1$ units)
\begin{gather}\label{Zeeman1}
H=\frac{\Delta}{2}\sigma_z+v(\eta p_x\sigma_x+p_y\sigma_y)+\left(
                                                            \begin{array}{cc}
                                                              0 & \mu p_+^2 \\
                                                              \mu p_-^2 & 0 \\
                                                            \end{array}
                                                          \right)
+s\eta\frac{\lambda_c}{2}(\sigma_z+1)
-s\eta\frac{\lambda_v}{2}(\sigma_z-1)+s\Delta_Z,
\end{gather}
where $\Delta$ is a bandgap, $\sigma_i$ are the Pauli matrices, $v$ is the band parameter with the dimensionality of velocity, $\eta=\pm1$ is the valley index, $\mathbf{p}$ is the electron momentum,  $p_{\pm}=\eta p_x\pm i p_y$, $\mu$ is the band parameter describing the trigonal warping and nonparabolicity of electron dispersion, $s$ is the $z$-component of electron spin, $\lambda_{c,v}$ describe spin-orbit splitting of the conduction and valence bands, and $\Delta_Z\propto B$ is the Zeeman energy. Eigenvalues of this Hamiltonian read 
\begin{gather}\label{Zeeman2}
E_{s\eta}({\bf p})=s\Delta_Z+s\eta\frac{\lambda_c+\lambda_v}{2}\pm\sqrt{\left[\frac{\Delta-s\eta(\lambda_v-\lambda_c)}{2}\right]^2+|h_{\bf p}|^2}\approx\frac{\Delta}{2}+s(\Delta_Z+\eta\lambda_c)+\frac{|h_{\bf p}|^2}{\Delta},\\\nonumber
\frac{|h_{\bf p}|^2}{\Delta}=\frac{p^2}{2m}+\eta w(p_x^3-3p_xp_y^2) +\frac{\mu^2}{\Delta}p^4,
\end{gather}
where $m=\Delta/2v^2$ is an effective mass (it equals half of free-electron mass, $m=0.5 m_0$ in MoS$_2$), $w=2v\mu/\Delta$ is a warping amplitude. Here, we consider only the conduction band (``$+$'' sign in Eq.(\ref{Zeeman2})) assuming high doping.

Let us now derive the formula for the photogalvanic effect (PGE) current density. 
The Boltzmann equation for the electron distribution function $f$ reads
\begin{gather}\label{Zeeman3}
\frac{\partial f}{\partial t}+e{\bf E}(t)\cdot\frac{\partial f}{\partial {\bf p}}=-\frac{f-f_0}{\tau};\,\,\,\,\,
{\bf E}(t)={\bf E}e^{-i\omega t}+{\bf E}^*e^{i\omega t},
\end{gather}
where $f_0$ is the Fermi distribution function, $\tau$ is a relaxation time; $\textbf{E}$ is the amplitude of the uniform electromagnetic (EM) field, which is perpendicular to the monolayer, thus, $\textbf{E}=(E_x, E_y, 0)$.

The following step is to expand the distribution function with respect to electric field as a perturbation, $f=f_0+f_1(t)+f_2+...$.
For the first-order contribution, it gives
\begin{gather}\label{Zeeman4}
f_1=-e\tau_{\omega}{\bf E}\cdot\frac{\partial f_0}{\partial {\bf p}}=e\tau_{\omega}{\bf v}\cdot {\bf E}(-f_0')\quad,\quad f^*_1=e\tau^*_{\omega}{\bf v}\cdot {\bf E}^*(-f_0'),
\end{gather}
where $\tau_\omega=\tau/(1-i\omega\tau)$, $\textbf{v}=\partial_{\textbf{p}}E_{s\eta}(\textbf{p})$, and $f_0'=\partial f_0/\partial E_{s\eta}$.

The second-order correction reads
\begin{gather}\label{Zeeman5}
e{\bf E}\frac{\partial f_1^*}{\partial {\bf p}}+e{\bf E}^*\frac{\partial f_1}{\partial {\bf p}}=-\frac{f_2}{\tau}\quad\Rightarrow\quad
f_2=-e\tau\left({\bf E}\frac{\partial f_1^*}{\partial {\bf p}}+{\bf E}^*\frac{\partial f_1}{\partial {\bf p}}\right),
\end{gather}
which gives the PGE current in form
\begin{gather}\label{Zeeman6}
j_\alpha=e\int\frac{d{\bf p}}{(2\pi)^2}v_\alpha\,f_2.
\end{gather}
Substituting here the corrections Eqs.~\eqref{Zeeman4} and~\eqref{Zeeman5}, and integrating by parts twice yields
\begin{gather}\label{Zeeman7}
j_\alpha=e^3\tau
\left(\tau^*_\omega E_\beta E^*_\gamma+\tau_\omega E^*_\beta E_\gamma\right)
\sum_{s,\eta}\int\frac{d{\bf p}}{(2\pi)^2}\frac{\partial^2 v_\alpha}{\partial p_\beta\partial p_\gamma}\,f_0[E_{s\eta}({\bf p})];\,\,\,\,\,
\frac{\partial^2 v_\alpha}{\partial p_\beta\partial p_\gamma}\equiv\frac{\partial^3 E_{s\eta}({\bf p})}{\partial p_\beta\partial p_\gamma\partial p_\alpha}.
\end{gather}
Here, $f_0[E_{s\eta}({\bf p})]=\theta[\epsilon_F-E_{s\eta}({\bf p})]$ for a degenerate electron gas. Presenting $\partial^2v_{\alpha}/\partial p_{\beta}\partial p_{\gamma}$ in explicit form, one can write the PGE current as $\textbf{\textit{j}}=\zeta_N \textbf{F}$, where $\textbf{F}=(|E_x|^2-|E_y|^2,-E_xE_y^{\ast}-E_yE_x^{\ast})$, and
\begin{eqnarray}\label{Zeeman8}
\zeta_N=12  \frac{e^3w\tau^2}{1+\omega^2\tau^2}
\sum_{s,\eta}\int\frac{d{\bf p}}{(2\pi)^2}\,\eta\, f_0[E_{s\eta}({\bf p})].
\end{eqnarray}
Eq.~(\ref{Zeeman8}) is proportional to the differences between numbers of electrons in both valleys. Thus, we conclude that the normal state electrons does not carry the non-reciprocal PGE response, as was claimed in \cite{nagaosa}. This is because contribution from the different bands cancel out each other in the lowest order over the trigonal warping amplitude and Zeeman energy. 

%
%
%
%

To obtain the finite PGE response, we modify an original model by considering energy-dependent relaxation time, $\tau_{\varepsilon}$, which may occur due to scattering on the Coulomb impurities. Thus, the Eq.~(\ref{Zeeman6}) may be rewritten as
\begin{gather}\label{Zeeman9}
j_\alpha=2e^3 E_\beta E^*_\gamma
\sum_{s,\eta}\int\frac{d{\bf p}}{(2\pi)^2}v_{\alpha}\tau_{\varepsilon}\frac{\partial}{\partial p_{\beta}}\left\{\tau_{\varepsilon}\frac{\partial}{\partial p_{\gamma}} f_0\right\}.
\end{gather}
Here, we consider only the static limit for simplicity, $\omega\tau_{\varepsilon}\ll 1$. 

Expanding the velocity, $v_{\alpha}=\partial E_{s\eta}(\textbf{p})/\partial p_{\alpha}=v_{\alpha}^0+\partial w_{\textbf{p}}/\partial p_{\alpha}$ (here, we denote $v_{\alpha}^0=p_{\alpha}/m$ and $w_{\textbf{p}}=\eta w (p_x^3-3p_xp_y^2)$ in accordance with the Eq.~(\ref{Zeeman2})), and relaxation time, $\tau_{\varepsilon}=\tau^{0}_{\varepsilon}+w_{\textbf{p}}(\tau^0_{\varepsilon})'$ (here, "$...'$" means the derivative over $\varepsilon^0_{\textbf{p}}=\textbf{p}^2/2m$), up to the first order in warping, and considering only $x$ component of the PGE tensor, we obtain few contribution into the current Eq.~(\ref{Zeeman9})
\begin{gather}
j_x^{(1)}=2e^3 |E_x|^2
\sum_{s,\eta}\int\frac{d{\bf p}}{(2\pi)^2}v_{x}^0\tau^0_{\varepsilon}\frac{\partial}{\partial p_{x}}\left\{v_x^0 w_{\textbf{p}}(\tau^0_{\varepsilon}f_0)'\right\},\label{eq1}\\
j_x^{(2)}=2e^3 |E_x|^2
\sum_{s,\eta}\int\frac{d{\bf p}}{(2\pi)^2}v_{x}^0\tau^0_{\varepsilon}\frac{\partial}{\partial p_{x}}\left\{\tau^0_{\varepsilon}f_0'\frac{\partial}{\partial p_{x}}w_{\textbf{p}}\right\},\label{eq2}\\
j_x^{(3)}=2e^3 |E_x|^2
\sum_{s,\eta}\int\frac{d{\bf p}}{(2\pi)^2}\tau^0_{\varepsilon}\frac{\partial w_{\textbf{p}}}{\partial p_{x}}\frac{\partial}{\partial p_{x}}\left\{\tau^0_{\varepsilon}v_{x}^0f_0'\right\},\label{eq3}\\
j_x^{(4)}=2e^3 |E_x|^2
\sum_{s,\eta}\int\frac{d{\bf p}}{(2\pi)^2}(\tau^0_{\varepsilon})'v_x^0w_{\textbf{p}}\frac{\partial}{\partial p_{x}}\left\{\tau^0_{\varepsilon}v_{x}^0f_0'\right\}.\label{eq4}
\end{gather}
Thus, integrating by parts in Eqs.~(\ref{eq1}) and(\ref{eq2}), and performing the integration over the polar angle, we obtain

\begin{gather}
j_x^{(1)}=-\frac{2e^3 |E_x|^2}{(2\pi)^2}
2 m\pi w\sum_{s,\eta}\eta \int_0^{\infty}d\varepsilon_{\textbf{p}}\varepsilon_{\textbf{p}}^3(\tau^0_{\varepsilon})'(\tau^0_{\varepsilon}f_0')',\label{eqq1}\\
j_x^{(2)}=-\frac{2e^3 |E_x|^2}{(2\pi)^2}
6 m\pi w\sum_{s,\eta}\eta \int_0^{\infty}d\varepsilon_{\textbf{p}}\varepsilon_{\textbf{p}}^2\tau^0_{\varepsilon}(\tau^0_{\varepsilon})'f_0',\\
j_x^{(3)}=-\frac{2e^3 |E_x|^2}{(2\pi)^2}
6 m\pi w\sum_{s,\eta}\eta \int_0^{\infty}d\varepsilon_{\textbf{p}}\varepsilon_{\textbf{p}}^2\tau^0_{\varepsilon}(\tau^0_{\varepsilon}f_0')'.\label{eqq3}
\end{gather}
$j_x^{(4)}$ does not contribute into the PGE current since the integration over the polar angle gives zero. Combining Eqs.~(\ref{eqq1})-(\ref{eqq3}), and assuming $\tau^0_{\varepsilon}=\tau_0\varepsilon_{\textbf{p}}$, we have
\begin{gather}\label{Zeeman10}
j_x=\frac{2e^3|E_x|^2}{(2\pi)^2}54w\pi m\sum_{s,\eta}\eta \tau_0^2 \int_0^{\infty}d\varepsilon_{\textbf{p}} \theta[\epsilon_F-s(\Delta_Z+\eta \lambda_c)-\varepsilon_{\textbf{p}}]\varepsilon_{\textbf{p}}^2=216\frac{e^3|E_x|^2}{\pi }\Delta_Z\lambda_c w m \tau_0^2\epsilon_F.
\end{gather}
Generalizing Eq.~(\ref{Zeeman10}) for all possible components of the PGE tensor, finally, we have (restoring dimensionality)
\begin{gather}\label{Zeeman10}
\textbf{\textit{j}}=108\frac{e^3\Delta_Z\lambda_c w  \tau_0^2 n_e}{ \hbar^3}\textbf{F}(\textbf{E}),
\end{gather}
where it was taken into account that $k_F=\sqrt{\pi n_e}$ in MoS$_2$, and $n_e$ is a concentration of electrons.

\vspace{20pt}
\section{Superconducting fluctuations contribution to the PGE}

\subsection{Ginzburg-Landau equation for the free energy}

The Ginzburg-Landau (GL) free energy (in terms of the order parameter) reads as~\cite{nagaosa}
\begin{gather}\label{GL}
F=\sum_{\textbf{q}}\left[\frac{1}{g}+T\sum_{n,k,\eta=\pm}G^e_{k+q,\uparrow,\eta}G^h_{k,\downarrow,-\eta}\right]|\Delta_{\textbf{q}}|^2,
\end{gather}
where $g$ is the strength of the attractive interaction, and
\begin{gather}
G^e_{ks\eta}=\frac{1}{i\omega_n-E_{s\eta}(k)}=\frac{1}{i\omega_n-\frac{k^2}{2m}-\eta w(k_x^3-3k_xk_y^2)+\epsilon_F-s(\Delta_Z+\eta \lambda_c)},\\
G^h_{ks\eta}=\frac{1}{i\omega_n+E_{s\eta}(-k)}=\frac{1}{i\omega_n+\frac{k^2}{2m}-\eta w(k_x^3-3k_xk_y^2)-\epsilon_F+s(\Delta_Z+\eta \lambda_c)}
\end{gather}
are the electron and hole Green's functions in Matsubara's frequency representation, $\omega_n=\pi T (2n+1)$.  
The Green's functions without the account of the trigonal warping read as $g^{e(h)}_k$. 
Note, that the GL equation is written for the s-wave pairing (see indexes of the Green's function in Eq.~(\ref{GL})).

The product of the Green's functions in  Eq.~(\ref{GL}) can be further expanded over small $q$ up to the cubic terms. 
The zeroth-order term is $2\nu \epsilon$, where $\nu=m/2\pi$ is the density of states in 2DEG, and $\epsilon=(T-T_c)/T_c$. The first-order term in $q$ is zero.

The second-order term reads
\begin{eqnarray}\label{second}
&&\frac{T}{2}\sum_{nk\eta}G^h_{k,\downarrow,-\eta}(\textbf{q}\cdot \nabla)^2G^e_{k,\uparrow\eta}=\frac{T}{8\pi^2}\sum_{n\eta}\int d\theta \int\frac{dk^2}{2}\frac{q^2k^2\cos^2\theta}{m^2}(g^e_k)^2(g^h_k)^2\nonumber\\
&&=q^2\frac{T}{8}\sum_{\eta}(\epsilon_F+\eta \lambda_c)\sum_n\frac{1}{|\omega_n|^3}=\frac{q^2}{4m}\frac{\nu}{(\pi T)^2}\sum_{\eta}(\epsilon_F+\eta\lambda_c)\frac{7\zeta(3)}{4}=\frac{q^2}{4m}\frac{7\zeta(3)}{2}\frac{\nu\epsilon_F}{(\pi T)^2}.
\end{eqnarray}

The third-order term is
\begin{eqnarray}\label{third}
&&\frac{T}{6}\sum_{nk\eta}G^h_{k,\downarrow,-\eta}(\textbf{q}\cdot \nabla)^3G^e_{k,\uparrow\eta}=\sum_{n\eta k}\eta T w f_{q}g^h_k(g^e_k)^2\left[1+6g^e_k\frac{k^2}{2m}+9(g^e_k)^2\left(\frac{k^2}{2m}\right)^2+\left(\frac{k^2}{2m}\right)^3(g^e_k)^2(4g^e_k-g^h_k)\right]\nonumber\\
&&\approx \sum_{\eta}\eta \pi T w f_q \Delta_Z \nu\sum_n\frac{3(\epsilon_F+\eta \lambda_c)^2}{|\omega_n|^5}=\sum_{\eta}\eta \pi T w f_q \Delta_Z \nu (\epsilon_F+\eta \lambda_c)^2\frac{93\zeta(5)}{16(\pi T)^5}= w f_q \Delta_Z \lambda_c (\nu\epsilon_F) \frac{93\zeta(5)}{4(\pi T)^4},
\end{eqnarray}
where $f_q=q_x^3-3q_xq_y^2$, and to derive Eq.~(\ref{third}), we expanded the Green's functions up to the first-order in $w$ and $\Delta_Z$.

Taking into account that $\Psi_q=\sqrt{\frac{7\zeta(3)\nu\epsilon_F}{2(\pi T_c)^2}}\Delta_q$, the free energy reads
\begin{gather}\label{freeenergy}
F=\sum_q\left[a+\frac{q^2}{4m}+\Lambda  (q_x^3-3q_xq_y^2)\right]|\Psi_q|^2+\frac{b}{2}|\Psi_q|^4,
\end{gather}
where $a$ and $b$ are conventional parameters of the GL theory, and
\begin{gather}
\Lambda=\frac{93\zeta(5)\Delta_Z \lambda_c w}{28\zeta(3)(\pi T_c)^2}.
\end{gather}

Furthermore, Eq.~(\ref{freeenergy}) can be generalized on the case of superconductor with impurities (which corresponds to the limit of a ``dirty'' superconductors, see \cite{nagaosa2}).
A finite relaxation time $\tau$ in the expressions for the Green's functions allows to phenomenologically account for the impurities.
Thus, the second-order term in the product of two Green's functions (see Eq.~(\ref{second})) reads
\begin{gather}
\frac{T}{2}\sum_{nk\eta}G^h_{k,\downarrow,-\eta}(\textbf{q}\cdot \nabla)^2G^e_{k,\uparrow\eta}=\frac{q^2}{4m}\frac{2\nu\epsilon_F }{(\pi T)^2}\sum_{n}\frac{1}{|2n+1|^2}\frac{1}{|2n+1|+\frac{1}{2T\tau}}
\nonumber\\
=\frac{q^2}{4m}\frac{2n_e\lambda_c}{(\pi T)^2}(2\pi T\tau)\left\{\psi\left(\frac{1}{2}\right)-\psi\left(\frac{1}{2}+\frac{1}{4\pi T\tau}\right)+\frac{1}{4\pi T\tau}\psi'\left(\frac{1}{2}\right)\right\}\nonumber\\
=\frac{q^2}{m}n_e\frac{4}{v_F^2}\xi^2 =q^2 \xi^2 \tilde \nu,
\end{gather}
where $n_e$ is a density of electrons, $\tilde\nu=m/\pi$, $v_F=\sqrt{4\pi n_e}/m$, and $\xi$ is the SC coherence length. 

The third-order term in the product of two Green's functions reads
\begin{gather}
\frac{T}{6}\sum_{nk\eta}G^h_{k,\downarrow,-\eta}(\textbf{q}\cdot \nabla)^3G^e_{k,\uparrow\eta}= \sum_{\eta}\eta \pi T w f_q \Delta_Z \nu\sum_n\frac{3(\epsilon_F+\eta \lambda_c)^2}{2|\omega_n|^2}\frac{1}{\left(|\omega_n|+\frac{1}{2\tau}\right)^2}\left(\frac{1}{|\omega_n|}+\frac{1}{|\omega_n|+\frac{1}{2\tau}}\right)\nonumber\\
=\frac{12wf_q\Delta_Z\lambda_c }{(\pi T)^4}\frac{-2\pi^2+4\psi^{(1)}\left(\frac{1}{2}+\frac{1}{4\pi T\tau}\right)+\frac{1}{2\pi T\tau}\left[14 \zeta(3)-\psi^{(2)}\left(\frac{1}{2}+\frac{1}{4\pi T\tau}\right)\right]}{16(2\pi T\tau)^{-3}}.
\end{gather}
Taking into account the well-known correspondence between the order parameter and SC wave function \cite{Varlamov}, $\tilde\nu \xi^2\Delta_q^2=(1/4m)\Psi_q^2$, Eq.~\eqref{freeenergy} slightly changes, transforming into
\begin{eqnarray}\label{freeenergy2}
F=\sum_q\left[a+\frac{q^2}{4m}+\Lambda_{\tau}  (q_x^3-3q_xq_y^2)\right]|\Psi_q|^2+\frac{b}{2}|\Psi_q|^4,
\end{eqnarray}
where $\Lambda_{\tau}=\Lambda\cdot f_{\tau}(2\pi T\tau)$ and 
\begin{eqnarray}\label{dirty_coeff}
f_{\tau}(x)=\frac{7\zeta(3)}{31\zeta(5)}\frac{\pi n_e}{m^2}\frac{x^3}{(\pi T \xi)^2 }\left\{-2\pi^2+4\psi'\left(\frac{1}{2}+\frac{1}{2x}\right)+\frac{1}{x}\left[14\zeta(3)-\psi''\left(\frac{1}{2}+\frac{1}{2x}\right)\right]\right\}.
\end{eqnarray}
%


\subsection{Time-dependent GL equation}
The time-dependent GL (TDGL) equation allows us to account for the effects of superconducting fluctuations (SFs).
Using Eq.~(\ref{freeenergy}), the TDGL equation with the scalar potential reads as
\begin{gather}\label{TDGL}
\left[
\gamma\frac{\partial}{\partial t}
+\varepsilon(\hat{\textbf{p}})+2ie\gamma\varphi(\textbf{r},t)
\right]
\Psi(\textbf{r},t)=f(\textbf{r},t),
\end{gather}
where $\varepsilon({\bf p})=p^2/4m+\alpha T_c\epsilon+\Lambda(p_x^3-3p_xp_y^2)=\alpha T_c(\epsilon +\xi^2p^2)+\Lambda(p_x^3-3p_xp_y^2)$ is the Cooper pair kinetic energy accounting for the trigonal warping, and $\alpha$ is the parameter of GL theory which connects an effective mass, $m$, and the coherence length via relation $4m\alpha T_c\xi^2=1$ (we use $\hbar=1$ units), which fixes the parameter $\alpha$ of the Ginzburg-Landau theory; $\epsilon=(T-T_c)/T_c$ is the reduced temperature, $\varphi$ is the scalar potential of the electric field, $f(\textbf{r},t)$ is the Langevin force describing the SFs in equilibrium. Parameters $\alpha$ and $\gamma$ are connected as $\gamma=\pi \alpha/8$.
The coherence length $\xi$ in 2D samples reads
\begin{gather}\label{EQ11}
\xi^2=\frac{v_F^2\tau^2}{2}\Bigl[
\psi\left(\frac{1}{2}\right)
-
\psi\left(\frac{1}{2}+\frac{1}{4\pi T\tau}\right)
+\frac{\psi'\left(\frac{1}{2}\right)}{4\pi T\tau}\Bigr],
\end{gather}
where $\psi(x)$ is the digamma function and $v_F$ is the Fermi velocity.

Scalar potential of the electric field (which corresponds to the incident uniform EM field) reads 
\begin{gather}\label{field}
\varphi(\textbf{r}, t)=\varphi e^{i\textbf{k} \textbf{r} -i\omega t}+\varphi^{\ast}e^{-i\textbf{k} \textbf{r} +i\omega t}.
\end{gather}
%
%
%
Using the expression for the quantum-mechanical operator of electric current, the current of the the Cooper pairs reads as
\begin{eqnarray}\label{current}
j&&=-\frac{i e\hbar}{2m}\left\{\Psi^{\ast}\nabla\Psi-\Psi\nabla \Psi^{\ast}\right\}=\frac{e}{2m}\left\{\Psi^{\ast}(-i\hbar\nabla)\Psi+\Psi(i\hbar\nabla )\Psi^{\ast}\right\}\nonumber\\
&&\Rightarrow j=\frac{2e}{2}\left\{\Psi^{\ast}v(\hat{ \textbf{p}})\Psi+\Psi v(-\hat{ \textbf{p}})\Psi^{\ast}\right\},
\end{eqnarray}
where $v(\hat{\textbf{p}})$ is a velocity operator for the Cooper pair. 


\subsection{The order parameter and PGE current}

Let us expand the order parameter $\Psi(\textbf{r},t)$ with respect to the scalar potential, $\Psi(x)=\Psi_0(x)+\Psi_1(x)+\Psi_2(x)+...$, where $x=(\textbf{r},t)$. 
The TDGL equation (\ref{TDGL}) for the $n$-th order reads 
\begin{gather}
\left[\gamma\frac{\partial}{\partial t}+2ie\gamma\varphi(x)+\varepsilon(\hat{\textbf{p}})\right]\left[\Psi_0(x)+\Psi_1(x)+\Psi_2(x)+...\right]=f(x),
\end{gather}
where 
\begin{gather}
 \varepsilon(\textbf{p})=\frac{\textbf{p}^2}{4m}+\alpha T_c\epsilon+\Lambda(p_x^3-3p_xp_y^2)=\alpha T_c\left(\epsilon+\xi^2\textbf{p}^2\right)+\Lambda(p_x^3-3p_xp_y^2).
\end{gather}
Extracting the corresponding order of smallness yields
\begin{gather}\label{TDGLcorrections}
\left[\gamma\frac{\partial}{\partial t}+\varepsilon(\hat{\textbf{p}})\right]\Psi_{0}(x)=f(x)
\quad\Rightarrow\quad
\Psi_{0}(x)=\int dx' g(x-x')f(x'),
\\\nonumber
\left[\gamma\frac{\partial}{\partial t}+\varepsilon(\hat{\textbf{p}})\right]\Psi_{1}(x)=-2ie\gamma \varphi(x)\Psi_{0}(x)
\quad\Rightarrow\quad
\Psi_{1}(x)=-2ie\gamma \int dx' g(x-x')\varphi(x')\Psi_{0}(x'),
\\\nonumber
\left[\gamma\frac{\partial}{\partial t}+\varepsilon(\hat{\textbf{p}})\right]\Psi_{2}(x)=-2ie\gamma \varphi(x)\Psi_{1}(x)
\quad\Rightarrow\quad
\Psi_{2}(x)=-2ie\gamma \int dx' g(x-x')\varphi(x')\Psi_{1}(x'),\nonumber\\
\Psi_{2}(x)=(-2ie\gamma)^2 \int dx_1\int dx_2 g(x-x_1)\varphi(x_1)g(x_1-x_2)\varphi(x_2)\Psi_{0}(x_2).\nonumber
\end{gather}
Here, $g(\textbf{r}, t)=\sum_{{\bf q},\varepsilon} e^{i\textbf{q}\textbf{r}-i\varepsilon t}g_{\textbf{q}}(\varepsilon)=\sum_{{\bf q},\varepsilon} e^{i\textbf{q}\textbf{r}-i\varepsilon t}[-i\gamma \varepsilon +\varepsilon(\textbf{q})]^{-1}$ is the fluctuation propogator.

Meanwhile, using Eq.~(\ref{current}), one can extract two second-order contributions to the current density:
\begin{gather}
j^{I}=\frac{(2e)}{2}\left\{\Psi_1^{\ast}v(\hat{\textbf{p}})\Psi_1+\Psi_1 v(-\hat{\textbf{p}})\Psi_1^{\ast}\right\},\\
j^{II}=\frac{(2e)}{2}\left\{\Psi_0^{\ast}v(\hat{\textbf{p}})\Psi_2+\Psi_2^{\ast}v(\hat{\textbf{p}})\Psi_0\right\}
+\frac{1}{2}\left\{\Psi_0 v(-\hat{\textbf{p}})\Psi_2^{\ast}+\Psi_2 v(-\hat{\textbf{p}})\Psi_0^{\ast}\right\}.
\end{gather}
%
%

\newpage
\textbf{I-type current density}

First, we find that 
\begin{gather}
\Psi_1(x)=-2ie\gamma \sum_{p\varepsilon}\int dx' g_{\textbf{p}}(\varepsilon)e^{ip(x-x')}\left(\varphi e^{ikx'}+\varphi^{\ast}e^{-ikx'}\right)\sum_{q\varepsilon'}e^{iqx'}g_{\textbf{q}}(\varepsilon')f_{\textbf{q}}(\varepsilon')\nonumber\\
=-2ie\gamma \sum_{q\varepsilon}g_{\textbf{q}}(\varepsilon)f_{\textbf{q}}(\varepsilon)\left(e^{i(k+q)x}g_{\textbf{k}+\textbf{q}}(\varepsilon+\omega)\varphi+e^{i(-k+q)x}g_{-\textbf{k}+\textbf{q}}(\varepsilon-\omega)\varphi^{\ast}\right),\\
\Psi_1^{\ast}(x)=2ie\gamma \sum_{q\varepsilon}g_{\textbf{q}}^{\ast}(\varepsilon)f_{\textbf{q}}^{\ast}(\varepsilon)\left(e^{-i(k+q)x}g_{\textbf{k}+\textbf{q}}^{\ast}(\varepsilon+\omega)\varphi^{\ast}+e^{-i(-k+q)x}g_{-\textbf{k}+\textbf{q}}^{\ast}(\varepsilon-\omega)\varphi\right).
\end{gather}

As a result,
\begin{eqnarray}
\Psi_1^{\ast}v(\textbf{p})\Psi_1&=&(2e\gamma)^2\sum_{\varepsilon\varepsilon'qq'}g^{\ast}_{\textbf{q}'}(\varepsilon')e^{-iq'x}\left[e^{-ikx}g^{\ast}_{\textbf{k}+\textbf{q}'}(\varepsilon'+\omega)\varphi^{\ast}+e^{ikx}g^{\ast}_{\textbf{-k+q}'}(\varepsilon'-\omega)\varphi\right]\nonumber\\
&&\times g_{\textbf{q}}(\varepsilon)\left[g_{\textbf{k+q}}(\varepsilon+\omega)v(\textbf{k}+\textbf{q})e^{i(k+q)x}\varphi+g^{\ast}_{\textbf{-k+q}}(\varepsilon-\omega)v(-\textbf{k}+\textbf{q})e^{i(q-k)x}\varphi^{\ast}\right]f_{\textbf{q}'}^{\ast}(\varepsilon')f_{\textbf{q}}^{\ast}(\varepsilon).
\end{eqnarray}
Taking into account that the averaging over fluctuating Langevin forces gives  $\langle f_{\textbf{q}}(\varepsilon)f_{\textbf{q}'}^{\ast}(\varepsilon')\rangle=2T {\rm Re}\{\gamma\}\delta_{\textbf{q}\textbf{q}'}\delta_{\varepsilon\varepsilon'}=f_0^2\delta_{\textbf{q}\textbf{q}'}\delta_{\varepsilon\varepsilon'}$, we find
\begin{gather}
\Psi_1^{\ast}v(\textbf{p})\Psi_1=(2e\gamma f_0)^2\sum_{\varepsilon q} |g_{\textbf{q}}(\varepsilon)|^2\left\{|g_{\textbf{k+q}}(\varepsilon+\omega)|^2v(\textbf{k}+\textbf{q})+|g_{\textbf{-k+q}}(\varepsilon-\omega)|^2 v(-\textbf{k}+\textbf{q})\right\}|\varphi|^2.
\end{gather}
Here, we consider only stationary components of the PGE current disregarding the second-order harmonics. In similar manner, one can get exactly the same expression for
\begin{gather}
\Psi_1v(-\textbf{p})\Psi_1^{\ast}=(2e\gamma f_0)^2\sum_{\varepsilon q} |g_{\textbf{q}}(\varepsilon)|^2\left\{|g_{\textbf{k+q}}(\varepsilon+\omega)|^2v(\textbf{k}+\textbf{q})+|g_{\textbf{-k+q}}(\varepsilon-\omega)|^2 v(-\textbf{k}+\textbf{q})\right\}|\varphi|^2.
\end{gather}

\vspace{20pt}
\textbf{II-type current density}

Performing a similar procedure, one finds
\begin{gather}
\Psi_0^{\ast}v(\textbf{p})\Psi_2=-(2e\gamma f_0)^2\sum_{\varepsilon q} |g_{\textbf{q}}(\varepsilon)|^2
\left\{g_{\textbf{q}}(\varepsilon)g_{\textbf{k}+\textbf{q}}(\varepsilon+\omega) +g_{\textbf{q}}(\varepsilon)g_{\textbf{q}-\textbf{k}}(\varepsilon-\omega) \right\}v(\textbf{q})|\varphi|^2,\\
\Psi_2^{\ast}v(\textbf{p})\Psi_0=-(2e\gamma f_0)^2\sum_{\varepsilon q} |g_{\textbf{q}}(\varepsilon)|^2
\left\{g_{\textbf{q}}^{\ast}(\varepsilon)g_{\textbf{k}+\textbf{q}}^{\ast}(\varepsilon+\omega) +g_{\textbf{q}}^{\ast}(\varepsilon)g_{\textbf{q}-\textbf{k}}^{\ast}(\varepsilon-\omega) \right\}v(\textbf{q})|\varphi|^2,\\
\Psi_0v(-\textbf{p})\Psi_2^{\ast}=-(2e\gamma f_0)^2\sum_{\varepsilon q} |g_{\textbf{q}}(\varepsilon)|^2
\left\{g_{\textbf{q}}(\varepsilon)g_{\textbf{k}+\textbf{q}}(\varepsilon+\omega) +g_{\textbf{q}}(\varepsilon)g_{\textbf{q}-\textbf{k}}(\varepsilon-\omega) \right\}v(\textbf{q})|\varphi|^2,\\
\Psi_2v(-\textbf{p})\Psi_0^{\ast}=-(2e\gamma f_0)^2\sum_{\varepsilon q} |g_{\textbf{q}}(\varepsilon)|^2
\left\{g_{\textbf{q}}^{\ast}(\varepsilon)g_{\textbf{k}+\textbf{q}}^{\ast}(\varepsilon+\omega) +g_{\textbf{q}}^{\ast}(\varepsilon)g_{\textbf{q}-\textbf{k}}^{\ast}(\varepsilon-\omega) \right\}v(\textbf{q})|\varphi|^2.
\end{gather}
%


\vspace{20pt}
\textbf{Total PGE current}

Combining all terms gives
\begin{gather}
j_{\alpha}=j_{\alpha}^{I}+j_{\alpha}^{II}=2e(2e\gamma f_0)^2\sum_{\textbf{q}\varepsilon }  |g_{\textbf{q}}(\varepsilon)|^2 \left\{|g_{\textbf{k+q}}(\varepsilon+\omega)|^2v_{\alpha}(\textbf{q}+\textbf{k})+|g_{\textbf{q}-\textbf{k}}(\varepsilon-\omega)|^2 v_{\alpha}(\textbf{q}-\textbf{k})\right\} |\varphi|^2\nonumber\\
- 2e(2e\gamma f_0)^2\sum_{\textbf{q}\varepsilon } |g_{\textbf{q}}(\varepsilon)|^2\left\{ g_{\textbf{q}}(\varepsilon)g_{\textbf{q}+\textbf{k}}(\varepsilon+\omega)+g_{\textbf{q}}(\varepsilon)g_{\textbf{q}-\textbf{k}}(\varepsilon-\omega)+g_{\textbf{q}}^{\ast}(\varepsilon)g_{\textbf{q}+\textbf{k}}^{\ast}(\varepsilon+\omega)+g_{\textbf{q}}^{\ast}(\varepsilon)g_{\textbf{q}-\textbf{k}}^{\ast}(\varepsilon-\omega)\right\}v_{\alpha}(\textbf{q})|\varphi|^2.
\end{gather}
It is more convenient to rewrite this expression in a different form:
\begin{gather}
j_{\alpha}=2e(2e\gamma f_0)^2\sum_{\textbf{q}\varepsilon}  |g_{\textbf{q}}(\varepsilon)|^2 \left\{|g_{\textbf{k+q}}(\varepsilon+\omega)|^2v_{\alpha}(\textbf{q}+\textbf{k})- v_{\alpha}(\textbf{q})[ g_{\textbf{q}}(\varepsilon)g_{\textbf{q}+\textbf{k}}(\varepsilon+\omega)+g_{\textbf{q}}^{\ast}(\varepsilon)g_{\textbf{q}+\textbf{k}}^{\ast}(\varepsilon+\omega)]\right\} |\varphi|^2  \nonumber\\
+(\textbf{k}\rightarrow -\textbf{k}, \omega \rightarrow -\omega).
\end{gather}
Performing the integration over the energy, we find
\begin{gather}
j_{\alpha}=\frac{2e(2e\gamma f_0)^2}{2\gamma}\sum_{\textbf{q}}   \left\{\frac{v_{\alpha}(\textbf{q}+\textbf{k})}{\varepsilon(\textbf{q})\varepsilon(\textbf{q}+\textbf{k})} -\frac{ v_{\alpha}(\textbf{q})}{\varepsilon^2(\textbf{q})}\right\} \frac{\varepsilon(\textbf{q})+\varepsilon(\textbf{q}+\textbf{k})}{\gamma^2\omega^2+[\varepsilon(\textbf{q})+\varepsilon(\textbf{q}+\textbf{k})]^2}|\varphi|^2  \nonumber\\
+(\textbf{k}\rightarrow -\textbf{k}, \omega \rightarrow -\omega).
\end{gather}
One can see that at $\textbf{k}\rightarrow 0$, expression above vanishes. Thus, one needs to expand the current density up to the second order over $\textbf{k}$ and use the correspondence between scalar potential and electric field to have $k_{\beta}k_{\gamma}|\varphi|^2=E_{\beta}E^{\ast}_{\gamma}$. 
The first-order expansion vanishes since there are always terms with opposite signs in front of $\textbf{k}$.

It is more convenient to rewrite equation above in the following form:
\begin{gather}
j_{\alpha}=\frac{2e(2e\gamma f_0)^2}{2\gamma}\sum_{\textbf{q}}   v_{\alpha}(\textbf{q})\left\{\frac{1}{\varepsilon(\textbf{q})\varepsilon(\textbf{q}-\textbf{k})} \frac{\varepsilon(\textbf{q})+\varepsilon(\textbf{q}-\textbf{k})}{\gamma^2\omega^2+[\varepsilon(\textbf{q})+\varepsilon(\textbf{q}-\textbf{k})]^2}-\frac{1}{\varepsilon^2(\textbf{q})}\frac{\varepsilon(\textbf{q})+\varepsilon(\textbf{q}+\textbf{k})}{\gamma^2\omega^2+[\varepsilon(\textbf{q})+\varepsilon(\textbf{q}+\textbf{k})]^2}\right\} |\varphi|^2  \nonumber\\
+({\bf k}\rightarrow -{\bf k}, \omega \rightarrow -\omega).
\end{gather}

Let us consider only $x$-component of the third-rank tensor ($j_x=\zeta_{xxx}E_xE_x^{\ast}$). 
Expanding $\varepsilon(\textbf{q}\pm\textbf{k})$ over small $\textbf{k}$ up to the second order,
we obtain
\begin{eqnarray}\label{fluc_current}
j_{x}=\frac{2e(2e\gamma f_0)^2}{\gamma}\frac{|E_x|^2}{4\pi^2}\int_0^{2\pi} d\theta\int_0^{\infty} dq q \frac{2}{4\varepsilon(\textbf{q})(\varepsilon^2(\textbf{q})+\frac{\gamma^2\omega^2}{4})}\left\{-v_x\frac{\partial v_x}{\partial q_x}\frac{1}{2\varepsilon(\textbf{q})}+v_x^3\frac{1}{2\varepsilon^2(\textbf{q})}+v_x^3\frac{4}{\gamma^2\omega^2+4\varepsilon^2(\textbf{q})}\right\}.
\end{eqnarray}

\vspace{20pt}
\textbf{Static limit}

In the static limit $\omega\rightarrow0$, let us expand $\varepsilon(\textbf{q})$ up to the first order over $\Lambda$ and using following notation $\varepsilon(\textbf{q})=\varepsilon_q^0+\Lambda f_q$. 
Then,
\begin{gather}
j_{x}=\frac{2e(2e\gamma f_0)^2}{\gamma}\frac{|E_x|^2}{8\pi^2}\int_0^{2\pi} d\theta\int_0^{\infty} dq^2 \frac{1}{4}\left\{-v_x\frac{\partial v_x}{\partial q_x}\frac{1}{(\varepsilon_{q}^{0})^4}\left(1-\frac{\Lambda f_q}{\varepsilon_q^{0}}\right)+v_x^3\frac{3}{(\varepsilon_{q}^0)^5}\left(1-\frac{5\Lambda f_q}{\varepsilon_q^0}\right)\right\}.
\end{gather}

After performing integration, we obtain
\begin{eqnarray}\label{static}
j_{x}=
\frac{2e(2e\gamma f_0)^2}{\gamma}\frac{|E_x|^2}{8\pi}\frac{\Lambda m}{(\alpha T_c \epsilon)^2}\frac{1}{4}(-8+18-6)=\frac{2e^3\pi\Lambda m}{64 T_c \epsilon^2}|E_x|^2.
\end{eqnarray}
%

Calculating PGE current for all components, we have $\textbf{\textit{j}}=(2e^3\pi \Lambda \pi m/64T_c \epsilon^2)\textbf{F}(\textbf{E})$, where $\textbf{F}(\textbf{E})=(|E_x|^2-|E_y|^2, -E_xE_y^{\ast}-E_x^{\ast}E_y)$, and
\begin{gather}
v_x=\frac{q_x}{2m}+3\Lambda (q_x^2-q_y^2) \quad,\quad v_y=\frac{q_y}{2m}-6\Lambda q_xq_y, \\ \frac{\partial v_x}{\partial q_x}=\frac{1}{2m}+6\Lambda q_x\quad,\quad \frac{\partial v_y}{\partial q_y}=\frac{1}{2m}-6\Lambda q_x\quad,\quad \frac{\partial v_x}{\partial q_y}=-6\Lambda q_y \\
v_x \frac{\partial v_x}{\partial q_x}\approx\frac{q_x}{4m^2}+\frac{3}{2}\frac{\Lambda}{m}(3q_x^2-q_y^2)\quad,\quad v_x^3\approx \frac{q_x^3}{8m^3}+\frac{9}{4}\frac{\Lambda}{m^2}(q_x^4-q_x^2q_y^2),\\
v_x\frac{\partial v_y}{\partial q_y}=\frac{q_x}{4m^2}-\frac{3}{2}\frac{\Lambda}{m}(q_x^2+q_y^2)\quad,\quad v_y\frac{\partial v_x}{\partial q_y}=-3\frac{\Lambda}{m}q_y^2\quad,\quad v_xv_y^2\approx\frac{q_xq_y^2}{8m^3}-\frac{3}{4}\frac{\Lambda}{m^2}(3q_x^2q_y^2+q_y^4).
\end{gather}

\vspace{20pt}
\textbf{Finite frequency limit}

For finite frequencies,  from Eq.~(\ref{fluc_current}) it follows that
\begin{eqnarray}
j_{x}=\frac{2e(2e\gamma f_0)^2}{\gamma}\frac{|E_x|^2}{8\pi^2}\frac{1}{4}\int_0^{2\pi} d\theta\int_0^{\infty} dq^2\left\{-v_x\frac{\partial v_x}{\partial q_x}\frac{1}{(\varepsilon_q^0)^2}\frac{1}{(\varepsilon_q^0)^2+\tilde\omega^2}\left(1-\frac{2\Lambda}{\varepsilon_q^0}f_q-\frac{2\varepsilon_q^0}{(\varepsilon_q^0)^2+\tilde\omega^2}\Lambda f_q\right)\right.\nonumber\\
+v_x^3\frac{1}{(\varepsilon_q^0)^3}\frac{1}{(\varepsilon_q^0)^2+\tilde\omega^2}\left(1-\frac{3\Lambda}{\varepsilon_q^0}f_q-\frac{2\varepsilon_q^0}{(\varepsilon_q^0)^2+\tilde\omega^2}\Lambda f_q\right)\nonumber\\
\left.+v_x^3\frac{1}{\varepsilon_q^0}\frac{1}{[(\varepsilon_q^0)^2+\tilde\omega^2]^2}\left(1-\frac{\Lambda}{\varepsilon_q^0}f_q-\frac{4\varepsilon_q^0}{(\varepsilon_q^0)^2+\tilde\omega^2}\Lambda f_q\right)\right\},
\end{eqnarray}
where $\tilde\omega=\gamma \omega/(2\alpha T_c)=\pi\omega/(16T_c)$.  
Performing integration, we find
\begin{gather}
j_{x}=\frac{2e^3\Lambda m \pi}{64 T_c}\frac{|E_x|^2}{4}\left\{-\frac{3\cdot 16}{2\tilde\omega^2}\left(\log\left(1+\frac{\tilde\omega^2}{\epsilon^2}\right)-2+\frac{\epsilon}{\tilde\omega}[\pi-2\arctan(\epsilon/\tilde\omega)]\right)\right.\nonumber\\
+\frac{9\cdot 8}{2\tilde\omega^2}\left[-1+\left(1+\frac{\epsilon^2}{\tilde\omega^2}\right)\log\left(1+\frac{\tilde\omega^2}{\epsilon^2}\right)\right]\nonumber\\
-\frac{32}{2\tilde\omega^2}\left[1-3\frac{\epsilon^2}{\tilde\omega^2}+3\frac{\epsilon}{\tilde\omega}\left(\frac{\epsilon^2-\tilde\omega^2}{\tilde\omega^2}\arctan\frac{\tilde\omega}{\epsilon}+\frac{\epsilon}{\tilde\omega}\log\left(1+\frac{\tilde\omega^2}{\epsilon^2}\right)\right)\right]\nonumber\\
-\frac{24}{6\tilde\omega^2}\left[-11+6\frac{\epsilon^2}{\tilde\omega^2}-6\frac{\epsilon}{\tilde\omega}\left(\frac{\epsilon^2-3\tilde\omega^2}{\tilde\omega^2}\right)\arctan\frac{\tilde\omega}{\epsilon}+\left(3-9\frac{\epsilon^2}{\tilde\omega^2}\right)\log\left(1+\frac{\tilde\omega^2}{\epsilon^2}\right)\right]\nonumber\\
-\frac{64}{16\tilde\omega^2}\left[4-3\pi\frac{\epsilon}{\tilde\omega}\left(1+\frac{\epsilon^2}{\tilde\omega^2}\right)+6\frac{\epsilon^2}{\tilde\omega^2}+6\frac{\epsilon}{\tilde\omega}\left(1+\frac{\epsilon^2}{\tilde\omega^2}\right)\arctan\frac{\epsilon}{\tilde\omega}\right],
\end{gather}
or after algebraic simplifications,
\begin{eqnarray}
j_{x}&=&\frac{2e^3\Lambda m \pi}{64 T_c}\frac{|E_x|^2}{\tilde\omega^2}6\left\{1+\log\left(1+\frac{\tilde\omega^2}{\epsilon^2}\right)\left(\frac{\epsilon^2}{\tilde\omega^2}\right)\right.\nonumber\\
&&\left.-\frac{\epsilon}{\tilde\omega}\arctan\frac{\tilde\omega}{\epsilon}\left(\frac{\epsilon^2}{\tilde\omega^2}+1\right)+\frac{\epsilon}{\tilde\omega}\arctan\frac{\epsilon}{\tilde\omega}\left(1-\frac{\epsilon^2}{\tilde\omega^2}\right)+\frac{\pi}{2} \frac{\epsilon}{\tilde\omega}\left(\frac{\epsilon^2}{\tilde\omega^2}-1\right)\right\}.
\end{eqnarray}

Generalization the PGE current for all components gives $\textbf{\textit{j}}=\zeta_S \textbf{F}(\textbf{E})$, where (after restoring dimensionality)
\begin{eqnarray}\label{result}
\zeta_S&&=\frac{2e^3\Lambda m \pi}{64 \hbar^3T_c\epsilon^2}\frac{6\epsilon^2}{\tilde\omega^2}\left\{1+\log\left(1+\frac{\tilde\omega^2}{\epsilon^2}\right)\left(\frac{\epsilon^2}{\tilde\omega^2}\right)\right.\nonumber\\
&&\left.-\frac{\epsilon}{\tilde\omega}\arctan\frac{\tilde\omega}{\epsilon}\left(\frac{\epsilon^2}{\tilde\omega^2}+1\right)+\frac{\epsilon}{\tilde\omega}\arctan\frac{\epsilon}{\tilde\omega}\left(1-\frac{\epsilon^2}{\tilde\omega^2}\right)+\frac{\pi}{2} \frac{\epsilon}{\tilde\omega}\left(\frac{\epsilon^2}{\tilde\omega^2}-1\right)\right\}.
\end{eqnarray}
This formula is applicable in the clean superconductor limit. 
The generalization on the case of a dirty superconductor can be performed using Eq~(\ref{freeenergy2}) instead of Eq.~(\ref{freeenergy}). 
As a result, $\Lambda$ in Eq.~(\ref{result}) should be replaced by $\Lambda_{\tau}=\Lambda\cdot f_{\tau}(\tau T)$, where $f_{\tau}$ obeys Eq.~(\ref{dirty_coeff}).

\newpage
\section{Kinetic equations description of SC fluctuations contribution}

Here, we re-derive  Eq.~(\ref{static}) using a slightly different technique based on the Boltzmann kinetic equation for the fluctuating Cooper pairs in the uniform electromagnetic field $\textbf{E}(t)=\textbf{E} e^{-i\omega t}+\textbf{E}^{\ast}e^{i\omega t}$,
\begin{eqnarray}
\frac{\partial f}{\partial t}+2e \textbf{E}\cdot\frac{\partial f}{\partial \textbf{p}}=-\frac{1}{\tau_{\textbf{p}}}(f-f_0),
\end{eqnarray}
where $f$ is the distribution function of the fluctuating Cooper pairs, $\tau_{\textbf{p}}=\gamma/(2\varepsilon_{\textbf{p}})$ is their effective lifetime, $\varepsilon_{\textbf{p}}=\alpha T_c (\epsilon+\xi^2\textbf{p}^2)+\Lambda (p_x^3-3p_xp_y^2)$, and $f_0=\langle |\Psi_{\textbf{p}}|^2\rangle=T/\varepsilon_{\textbf{p}}$. 

Let us expand the distribution function, $f=f_0+f_1(t)+f_1^{\ast}(t)+f_2+...$. 
As a result,
\begin{eqnarray}
i\omega f_1 +2e \textbf{E}\cdot\frac{\partial f_0}{\partial \textbf{p}}=-\frac{f_1}{\tau_{\textbf{p}}} \quad\Rightarrow\quad f_1=-(2e)\frac{\tau_{\textbf{p}}}{1-i\omega \tau_{\textbf{p}}} \textbf{E}\cdot\frac{\partial f_0}{\partial \textbf{p}} \quad,\quad f_1^{\ast}=-(2e)\frac{\tau_{\textbf{p}}}{1+i\omega \tau_{\textbf{p}}} \textbf{E}^{\ast}\cdot\frac{\partial f_0}{\partial \textbf{p}},\\
(2e)\left\{\textbf{E}\cdot\frac{\partial f_1^{\ast}}{\partial \textbf{p}}+\textbf{E}^{\ast}\cdot\frac{\partial f_1}{\partial \textbf{p}}\right\}=-\frac{f_2}{\tau_{\textbf{p}}}\quad\Rightarrow\quad f_2=(2e)^2\tau_{\textbf{p}}\left\{\textbf{E}\cdot\frac{\partial }{\partial \textbf{p}}\left[\frac{\tau_{\textbf{p}}}{1+i\omega \tau_{\textbf{p}}}\textbf{E}^{\ast}\cdot\frac{\partial f_0}{\partial \textbf{p}}\right]+c.c.\right\}.
\end{eqnarray}
%
The PGE current density reads as 
\begin{eqnarray}
j_{\alpha}=2e\int \frac{d\textbf{p}}{(2\pi)^2}v_{\alpha}f_2(\textbf{p}),
\end{eqnarray}
or using expression for $f_2$, 
\begin{eqnarray}
j_{\alpha}=(2e)^3\int \frac{d\textbf{p}}{(2\pi)^2}v_{\alpha}\tau_{\textbf{p}}\left\{\textbf{E}\cdot\frac{\partial }{\partial \textbf{p}}\left[\frac{\tau_{\textbf{p}}}{1+i\omega \tau_{\textbf{p}}}\textbf{E}^{\ast}\cdot\frac{\partial f_0}{\partial \textbf{p}}\right]+c.c.\right\}.
\end{eqnarray}

In the static case, $\omega\rightarrow 0$, and after integrating by part, the current density reads
\begin{eqnarray}
j_{\alpha}=-2(2e)^3\int \frac{d\textbf{p}}{(2\pi)^2}\textbf{E}\cdot\frac{\partial v_{\alpha}\tau_{\textbf{p}}}{\partial \textbf{p}}\tau_{\textbf{p}}\textbf{E}^{\ast}\cdot\frac{\partial f_0}{\partial \textbf{p}}.
\end{eqnarray}
After straightforward manipulations, the $x$ component of the conductivity tensor turns into
\begin{eqnarray}
\zeta_{xxx}&=&2(2e)^3\int \frac{d\textbf{p}}{(2\pi)^2}\frac{\partial v_{x}\tau_{\textbf{p}}}{\partial p_{x}}\tau_{\textbf{p}}v_{x}\left(-\frac{\partial f_0}{\partial \varepsilon_{\textbf{p}}}\right)=2(2e)^3\int \frac{d\textbf{p}}{(2\pi)^2}\tau_{\textbf{p}}v_{x}\frac{\partial v_{x}\tau_{x}}{\partial p_{x}}\frac{T}{\varepsilon^2_{\textbf{p}}}=(2e)^3\int \frac{d\textbf{p}}{(2\pi\hbar)^2}\frac{\partial (v_{x}\tau_{x})^2}{\partial p_{x}}\frac{T}{\varepsilon^2_{\textbf{p}}}\nonumber\\
&=&2(2e)^3\int \frac{d\textbf{p}}{(2\pi)^2} \tau_{x}^2v_x^3\frac{T}{\varepsilon^3_{\textbf{p}}}=2(2e)^3\int \frac{d\textbf{p}}{(2\pi)^2} \left(\frac{\gamma}{2\varepsilon_{\textbf{p}}}\right)^2v_x^3\frac{T}{\varepsilon^3_{\textbf{p}}}=\frac{(2e)^3 T \gamma^2}{2(2\pi )^2}\int d\textbf{p}  \frac{v_x^3}{\varepsilon^5_{\textbf{p}}}.
\end{eqnarray}
Now, we expand $\varepsilon_{\textbf{p}}=\varepsilon^0_{\textbf{p}}+\Lambda f_{\textbf{p}}$ over small trigonal warping up to the first order in $\Lambda$ and perform integration over the 2D momentum, yielding
\begin{eqnarray}
&&\zeta_{xxx}=\frac{(2e)^3 T \gamma^2}{2(2\pi )^2}\int_0^{2\pi} d\theta \int_0^{\infty} \frac{dp^2}{2}\left\{\frac{9}{4}\frac{\Lambda}{m^2}p^4\frac{\cos^4\theta-\sin^2\theta}{\varepsilon^5_{\textbf{p}}}-\frac{5}{8}\frac{\Lambda}{m^3}p^6\frac{\cos^6\theta-3\cos^4\theta\sin^2\theta}{\varepsilon^6_{\textbf{p}}}\right\}\nonumber\\
&&=\frac{(2e)^3 T \gamma^2}{2(2\pi )^2} \pi\int_0^{\infty} \frac{dy}{2}\left\{\frac{9}{8}\frac{\Lambda}{\xi^6m^2}\frac{y^2}{(\alpha T_c)^5(\epsilon+y)^5}-\frac{5}{32}\frac{\Lambda}{\xi^8m^3}\frac{y^3}{(\alpha T_c)^6(\epsilon+y)^6}\right\}\nonumber\\
&&=\frac{(2e)^3 T \gamma^2}{2(2\pi)^2} \frac{\pi\Lambda m}{(\alpha T_c \epsilon)^2}2 \quad\Rightarrow\quad \zeta_{S}=\frac{2e^3\pi\Lambda m}{64\hbar^3T_c \epsilon^2},
\end{eqnarray}
thus we restore Eq.~(\ref{static}), $j=\zeta_S|E_x|^2$.

\bibliography{biblio}
\bibliographystyle{apsrev4-2}
